\documentclass[aps,pra, amsmath,amssymb,preprint,superscriptaddress]{revtex4-1}
\usepackage{color,soul}
\usepackage{threeparttable}
\usepackage{graphicx}% Include figure files
\usepackage{bm}  % bold in math
\usepackage{amssymb}
\usepackage{amsmath}
\usepackage{varioref}
\usepackage[dvipsnames]{xcolor}
\usepackage[colorlinks=true, linkcolor=blue, citecolor=blue,urlcolor=blue,breaklinks=true]{hyperref}

\newcommand{\ket}[1]{\left|{#1}\right\rangle}
\newcommand{\bra}[1]{\left\langle{#1}\right|}

\newcommand{\beq}{\begin{equation}}
\newcommand{\eeq}{\end{equation}}

\begin{document}
\title{Efficient and pure femtosecond-pulse-length source of polarization-entangled photons}

\author{Morgan M. Weston}
\affiliation{Centre for Quantum Dynamics and Centre for Quantum Computation and Communication Technology, Griffith University, Brisbane, Queensland 4111, Australia}
\author{Helen M. Chrzanowski}
\affiliation{Centre for Quantum Dynamics and Centre for Quantum Computation and Communication Technology, Griffith University, Brisbane, Queensland 4111, Australia}
\affiliation{Clarendon Laboratory, University of Oxford, Parks Road, Oxford OX1 3PU, UK}
\author{Sabine Wollmann}
\affiliation{Centre for Quantum Dynamics and Centre for Quantum Computation and Communication Technology, Griffith University, Brisbane, Queensland 4111, Australia}
\author{Allen Boston}
\affiliation{Centre for Quantum Dynamics and Centre for Quantum Computation and Communication Technology, Griffith University, Brisbane, Queensland 4111, Australia}
\author{Joseph Ho}
\affiliation{Centre for Quantum Dynamics and Centre for Quantum Computation and Communication Technology, Griffith University, Brisbane, Queensland 4111, Australia}
\author{Lynden K. Shalm}
\affiliation{National Institute of Standards and Technology, 325 Broadway, Boulder, Colorado 80305, USA}
\author{Varun B. Verma}
\affiliation{National Institute of Standards and Technology, 325 Broadway, Boulder, Colorado 80305, USA}
\author{Michael S. Allman}
\affiliation{National Institute of Standards and Technology, 325 Broadway, Boulder, Colorado 80305, USA}
\author{Sae Woo Nam}
\affiliation{National Institute of Standards and Technology, 325 Broadway, Boulder, Colorado 80305, USA}
\author{Raj B. Patel}
\affiliation{Centre for Quantum Dynamics and Centre for Quantum Computation and Communication Technology, Griffith University, Brisbane, Queensland 4111, Australia}
\author{Sergei Slussarenko}
\affiliation{Centre for Quantum Dynamics and Centre for Quantum Computation and Communication Technology, Griffith University, Brisbane, Queensland 4111, Australia}
\author{Geoff J. Pryde }
\email{g.pryde@griffith.edu.au}
\affiliation{Centre for Quantum Dynamics and Centre for Quantum Computation and Communication Technology, Griffith University, Brisbane, Queensland 4111, Australia}

\begin{abstract}
We present a source of polarization entangled photon pairs based on spontaneous parametric downconversion engineered for frequency uncorrelated  telecom photon generation. Our source provides photon pairs that display, simultaneously, the key properties for high-performance quantum information and fundamental quantum science tasks. Specifically, the source provides for high heralding efficiency, high quantum state purity and high entangled state fidelity at the same time. Among different tests we apply to our source we observe almost perfect non-classical interference between photons from independent sources with a visibility of $(100\pm5)\%$. 
\end{abstract}

\maketitle

\section{Introduction}	

Optical quantum information science offers powerful techniques for secure messaging and networking, processing and metrology, with significant advantages over techniques based on the laws of classical physics alone~\cite{Gisin2002,ralph09,OBrien2009,Xiang2011,Banaszek2009}. Photons also play a major role in fundamental studies on the nature of our world~\cite{shalm15,giustina15,hensen15,kocsis15}. An essential component for many optical quantum techniques and technologies is an efficient source of high-quality single photons or entangled photon pairs.  The ultimate goal is a deterministic and on-demand photon source, of which development is ongoing~\cite{he13,ding16}. One approach to building a deterministic source is the networked switching of high-quality probabilistic sources~\cite{Migdall2002,Collins2013}, such as heralded photon pair sources based on spontaneous parametric downconversion (SPDC). In the short- to medium-term, heralded pair sources without switching remain the favored architecture for modern experimental quantum optics because of the excellent capabilities they provide.

In this work, we demonstrate a very high quality SPDC source using periodically-poled KTiOPO$_4$ (ppKPT) in a polarizing Sagnac interferometer geometry. In particular, we demonstrate very high nonclassical (Hong-Ou-Mandel) interference between telecom photons from different sources, while maintaining high heralding efficiency  (the  probability of photon detection once the other photon or the pair is detected). Our source can also be configured to produce high quality polarization entanglement between telecom photons (also with high heralding efficiency). It uses a femtosecond pump laser, providing advantages in cost and simplicity over picosecond-pulse-length sources. 

Our goals are motivated by the key requirements of photonic quantum information experiments. The photons should be of high purity, so to provide high non-classical interference visibility, which is a crucial component of most quantum state transformation circuits. High heralding efficiency is essential for many applications, including loophole-free entanglement certification protocols. Alongside brightness, high heralding efficiency is a crucial ingredient in improving the practical scalability of multi-photon experiments which would require multiple sources of this kind. These include protocols based on quantum teleportation and entanglement swapping, the core components of a future quantum repeater. Additionally, a precious resource that can be obtained from the pair source is entanglement in a specific degree of freedom (e.g. polarization). Finally, in multi-photon experiments, photon pairs need to be precisely localized in time, so the source relies on a femto- or pico-second pulse-length laser pump, instead of continuous-wave (CW) pumping. Although each of these features is achievable with SPDC, no optimal source configuration that demonstrates all of these properties has been presented so far. Here we report on a polarization-entangled SPDC source in the telecom $1570~\textrm{nm}$ wavelength range, which provides all of these properties at the same time.

For the last two decades, the workhorses of photonic quantum information and communication have been bulk crystal sources. The most common approach utilizes non-collinear emission from type-II bulk barium borate (BBO) crystals, where collection from intersecting downconversion cones yields a polarization-entangled state~\cite{Kwiat:1995ub}.
While these traditional sources are still widely used for many photonic quantum information implementations -  most notably those concerned with multi-photon experiments - the past decade has seen the development of a new generation of photon sources that exploit recent advances in nonlinear optics.  

Quasi-phase matching via periodic poling overcomes many of the limitations of  angle phase matched sources, which include lower heralding efficiencies and lack of control over spectral properties of downconverted photons. One can utilize comparatively long crystals in collinear geometries, yielding both an inherently higher spectral flux and higher collection efficiencies. 
However, unlike the traditional bulk-crystal sources in which polarization entanglement is engineered within the emission cones, an additional step  is required to polarization-entangle the emitted photons. 

In 2006, Kim {\it et al.} provided the first demonstration of polarization-entangled photons from a ppKTP crystal embedded within a polarizing Sagnac interferometer utilizing a CW pump~\cite{Kim:2006cy, Wong:2006gt}. It was further advanced by Fedrizzi~{\it et al.}, yielding a source that was simultaneously bright, highly entangled and widely spectrally tuneable~\cite{Fedrizzi:2007tl}. 
In the CW regime, Sagnac-based sources are rapidly surmounting traditional bulk sources, with the increased brightness and potential for high heralding efficiencies allowing access to new and more comprehensive quantum information demonstrations~\cite{Smith:2011cc,Bennet:2012ch,giustina15}. However, CW pumping is incompatible with multi-photon experiments and the Sagnac architecture was first extended to the pulsed regime with narrowband pump pulses~\cite{Kuzucu:2008io} and later in the non-degenerate case of femtosecond pump pulses ~\cite{scheidl14} .

However, expanding the reach of photonic quantum information demands more than brightness alone. Ambitious implementations concerned with interfering multiple downconversion sources are hindered by limitations in purity, indistinguishability and heralding efficiency. 
In traditional SPDC sources, bi-photons are born spectrally entangled. Photon-counting measurements insensitive to this spectral information degrade the purity and indistinguishability of the state. 
These spectral correlations are traditionally erased by harsh spectral filtering of the state at the cost of both brightness and heralding efficiency. A more sophisticated approach to the problem is to instead directly engineer the process of downconversion itself to generate inherently pure bi-photons through group velocity matching (GVM)~\cite{Grice:1997tk,Keller:1997wq}. 
GVM was first demonstrated in 2004 in the context of second harmonic generation~\cite{Konig:2004en} in ppKTP, and has since been been used to generate spectrally pure photons in bulk KDP~\cite{Mosley:2008ir,Jin:2011dt}, bulk~\cite{Evans:2010jn,Gerrits:11,Jin:2013gg,Bruno:2014wi} and waveguided ppKTP~\cite{Eckstein:2011gpa,Harder:13} and bulk BBO~\cite{Lutz:2013uu}. 
The direct engineering of both the pump and crystal parameters improves the spectral factorability of the source, which improves the brightness and heralding efficiency by virtue of avoiding extensive spectral filtering. 
The source that we report on unites the complementary ideas of brightness through quasi-phase matching and collinear geometries, and engineered spectral purity and degeneracy. It also amounts to the first demonstration of a Sagnac interferometer to generate polarization entanglement in the femtosecond regime. This work naturally divides into two parts. Following~\cite{bennink2010optimal}, we first consider the spectral engineering of the downconversion source to produce degenerate and factorable photon pairs at $1570~\textrm{nm}$. We then proceed to demonstrate polarization entanglement using a Sagnac interferometer. We characterize our source's performance with a number of tests, such as joint spectral intensity measurement, state characterization  through polarization tomography and two-source Hong-Ou-Mandel (HOM) interference.

\section{Polarization unentangled source design and optimization}
\begin{figure}[t]
\begin{center}
\includegraphics[width = 0.95 \textwidth]{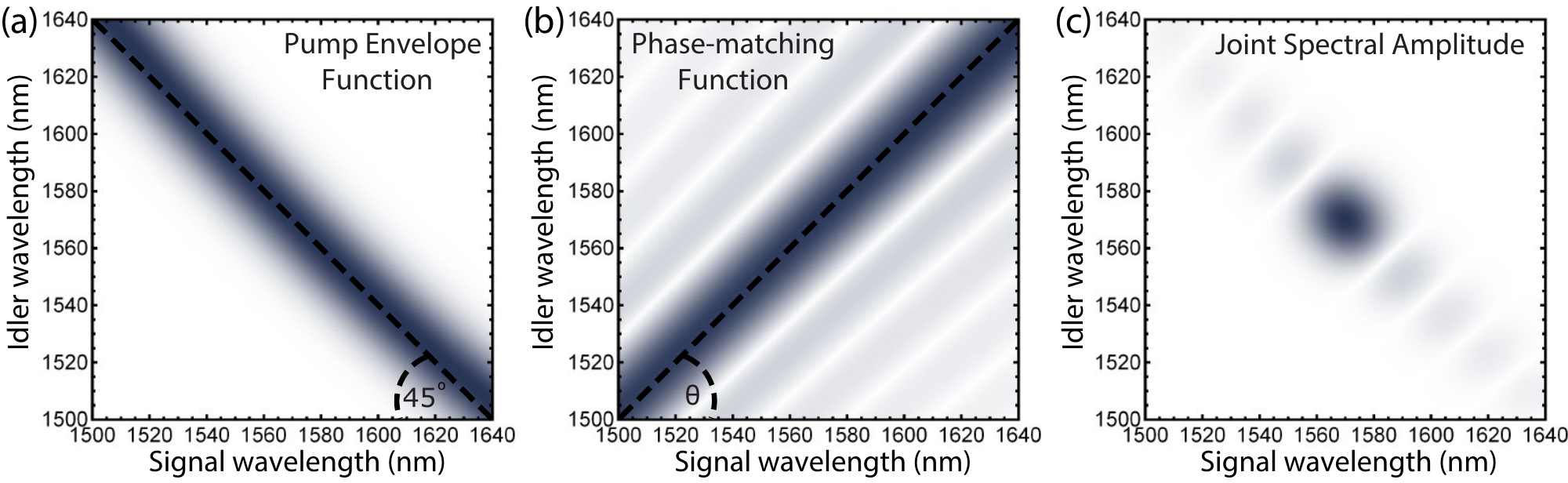}
\caption{The theoretical (a) pump envelope function, (b) phase-matching function, and corresponding (c) joint spectral amplitude for the following chosen source parameters: we consider type-II collinear downconversion with a pulsed pump field at $785~\textrm{nm}$ with a spectral bandwidth of $5.35~\textrm{nm}$, pumping a $2~\textrm{mm}$ ppKTP crystal with a poling period of $46.15~\mu\textnormal{m}$, phasematched to yield degenerate daughter photons at $1570~\textrm{nm}$. The joint spectral amplitude $f(\omega_s,\omega_i)$ is given by the product of the pump and phase functions intensities.
}
\label{theoryplots}
\end{center}
\end{figure}

SPDC is a $\chi  ^2$ non-linear process in which a pump photon of frequency $\omega_p$ is spontaneously split into a pair of daughter photons, by convention labeled the signal, $\omega_s$, and the idler, $\omega_i$, with their individual frequencies constrained by energy and momentum conservation. The two-photon wave function can be modeled as a integral over all possible frequency modes~\cite{bennink2010optimal},
\begin{eqnarray}
\left| \psi \right> = 
\int\displaylimits_{0}^\infty \int\displaylimits_{0}^\infty d\omega_s \, d\omega_i \; f(\omega_s,\omega_i) a^\dagger(\omega_s) a^\dagger(\omega_i)| 0, 0 \rangle  \; ,
\label{equ:SPDCStatebenn}
\end{eqnarray}	
where $a^\dagger(\omega_s)$ and $a^\dagger(\omega_i)$ are the creation operators of the signal $s$ and idler $i$ modes. The correlation function $f(\omega_s,\omega_i)$ represents the two-photon joint spectral amplitude, which can be decomposed as the product of its pump envelope function and the phase-matching function (see Fig.~\ref{theoryplots}) ,
\begin{eqnarray}
 f(\omega_s,\omega_i)= \mathcal{N} \, \alpha (\omega_s+\omega_i)\phi(\omega_s, \omega_i), \label{equ:JointAmp}
 \end{eqnarray}	
where $\mathcal{N}$ represents a normalization factor. The pump function is assumed to be Gaussian with a width proportional to the spectral bandwidth of the pump field. For a central frequency  $\omega_p$ and bandwidth $\sigma_p$, the pump function is given by
\begin{align}
\alpha (\omega_s+\omega_i)= \exp{\left[ - \dfrac{(\omega_s+\omega_i-\omega_p)^2}{\sigma_p^2}   \right]  } .
\label{equ:Pumpfunction}
\end{align}
In the collinear case, the phase matching function is a sinc function with a width inversely proportional to the length of the crystal and can be expressed as
\begin{align}
\phi(\omega_s, \omega_i)= \textrm{sinc} \left\lbrace  \dfrac{L}{2} \Delta K  \right\rbrace \nonumber \exp \left\lbrace -i\dfrac{L}{2} \Delta K  \right\rbrace.
\label{equ:phasefunction}
\end{align}
Here, $\Delta K= k_p(\omega_s+\omega_i)-k_s(\omega_s)-k_i(\omega_i)-\dfrac{2 \pi }{\Lambda}$ is phase mismatch, $k_i$,  $k_s$,  $k_p$ are the wave numbers of idler, signal and pump photons respectively and $\Lambda$ is the poling period. 
In this work, the joint spectral amplitude (JSA), pair collection probabilities, heralding ratios and spectral purity are theoretically modeled following the theoretical work of Bennink~\cite{bennink2010optimal} regarding collinear SPDC. Among various literature, the use of Sellmeier equations obtained from~\cite{fradkin1999tunable, konig2004extended} in combination with the thermal expansion described in~\cite{emanueli2003temperature} provided us the theoretical predictions that were in strong agreement with our experimental results. We also note that recently, a new tool for SPDC parameters calculation has appeared online~\cite{spdcalc}. 

To create a factorable JSA as shown in Fig.~\ref{theoryplots}(c) we must meet three conditions. Firstly, the peak contour of the phase matching function, defined by $ k_p(\omega_s+\omega_i)-k_s(\omega_s)-k_i(\omega_i)-\dfrac{2 \pi }{\Lambda} =0$,  must be aligned perpendicular to the fixed $45^\circ$ pump function (Fig.~\ref{theoryplots}(a)), therefore $\theta$ in Fig.~\ref{theoryplots}(b) must be $45^\circ$. Using the lowest order terms in the power expansion about the center frequencies and assuming perfect phase matching at the center frequencies $\overline{\omega_j}$, we are left with $0={k'}_p( \omega_s - \overline{\omega}_s )+{k'}_p( \omega_i - \overline{\omega}_i ) -{k'}_i( \omega_i - \overline{\omega}_i )-{k'}_s( \omega_s - \overline{\omega}_s ) $, $k'= \partial k /\partial \omega$, which can be written as 
\begin{align}
\dfrac{{k'}_s-{k'}_p}{{k'}_p-{k'}_i}=\dfrac{ \omega_i - \overline{\omega}_i  }{ \omega_s - \overline{\omega}_s }=\tan \theta. 
\end{align} 
For a circular JSA $\theta=45^\circ$ we obtain a group velocity matching condition of $k'_p=\dfrac{k'_s+k'_i}{2}$, where the group velocity is given by $\nu=\dfrac{1}{k'}$, which corresponds to an  optimal degenerate downconversion wavelength of $1582~\mathrm{nm}$ for our KTP crystal. For practicality we consider a center pump wavelength of $785~\mathrm{nm}$, resulting in a central downconversion wavelength of $1570~\mathrm{nm}$.  This deviation will result in a slightly elliptical JSA, however will not make a significant reduction to the state purity. Secondly, the crystal length and pump bandwidths must be matched. A $2~\textrm{mm}$ crystal length is chosen to match our Fourier-limited $\sim \! 170~\textrm{fs}$ laser pulses, the spectral purity maximized by a pump bandwidth of $5.35~\textrm{nm}$. Finally we then select a poling period to achieve phase matching at these wavelengths, with a period of $46.15~\mu\textrm{m}$ yielding degeneracy at $1570~\textrm{nm}$ at normal incidence and room temperature. 

We experimentally implemented this design in the laboratory. We used a modelocked Ti:Sapphire laser with a center wavelength of $785~\textrm{nm}$ to pump a ppKTP crystal in the collinear geometry and with type-II phasematching, with experimental parameters as listed above. We separated the (initially) co-propagating downconverted photons using a polarizing beam splitter and filtered away the pump beam using a dichroic mirror and a long pass filter ({\it Semrock FF875-Di01}~\cite{disclaimer} and {\it LP02-808RU}~\cite{disclaimer} , respectively). The photons were coupled into single-mode optical fibers with good mode matching, and were detected using superconducting nanowire single photon detectors (SNSPDs)~\cite{Marsili2013} with system quantum efficiencies $\sim 80-85\%$.   In order to compensate for the polarization-dependent behavior of SNSPDs, we optimized the quantum efficiency with fiber polarization controllers, when the photons that were coupled into fiber always had the same polarization (e.g. after a polarizing element). If a measurement across several polarizations was required, then either the FPCs were adjusted for each, or several random unitary were applied to the fibers (by looping the fibers differently over time) to average the polarization state. 

\begin{figure*}[t]
\begin{center}
\includegraphics[width = 0.95 \textwidth]{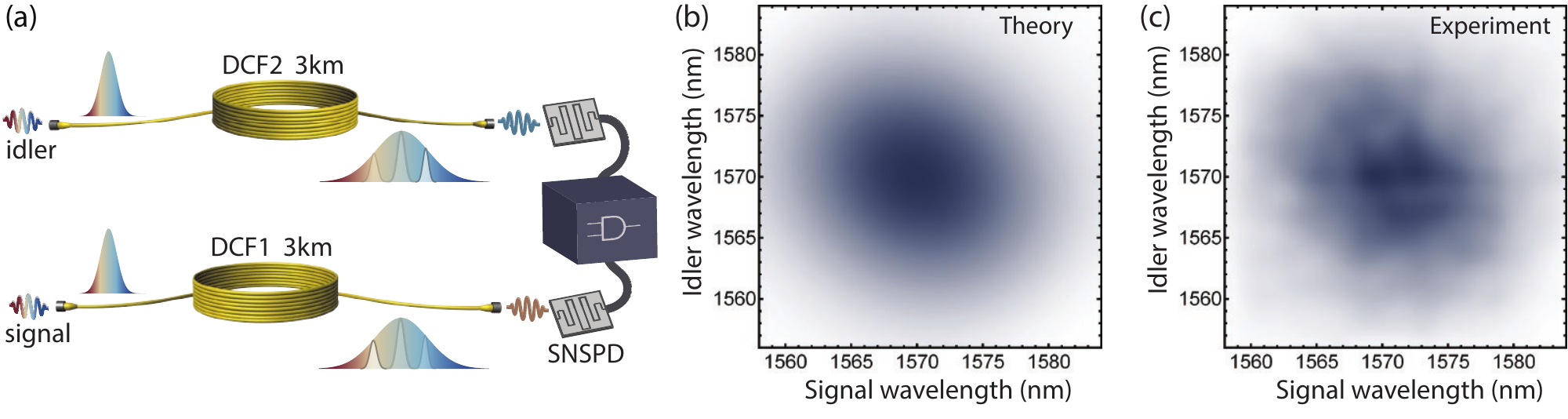}
\caption{Comparison of the (b) theoretically calculated and (c) experimentally measured joint spectral intensity, directly sampled using two fiber-based time resolved spectrometers, detailed in (a). The bandwidth of the spectrometer is limited by the $\approx12.3~\textrm{ns}$ temporal distance between pump pulses and is $\approx30~\textrm{nm}$.}
\label{theoryandspectra}
\end{center}
\end{figure*}

To confirm the spectral factorability of the downconverted photon pairs, the joint spectral intensity (JSI) was sampled via two time-of-flight spectrometers \cite{avenhaus2009fiber,Gerrits:11}. The group velocity dispersion (GVD) in dispersion compensation fibers (DCF) was used to temporally stretch the downconverted photons, providing sufficient temporal resolution to resolve the spectral correlations (or lack thereof) manifesting in arrival times between our SPDC pairs (see Fig.~\ref{theoryandspectra}(a)). 
The GVDs of our two DCFs were characterized by measuring the fiber delays with a tuneable laser in the telecom band. The spectrometer's resolution, limited by temporal distance between consecutive pump pulses and the time bin size of the coincidence logic card, are estimated to be $\approx 0.31$ and $0.33~\textrm{nm}$. Figures~\ref{theoryandspectra}(b) and~\ref{theoryandspectra}(c) show a comparison of our theoretically modeled JSI and our experimentally measured JSI. The results support the frequency degeneracy and factorability of the bi-photon state. We note that we also performed the characterisation using a crystal with a poling period of  $46.55~\mu\textnormal{m}$, as reported in~\cite{Gerrits:11}. In this case, we were able to experimentally achieve a degenerate spectrum only when the crystal tilt was adjusted away from the normal incidence, an observation which was also in agreement with our theoretical calculations. 

The optimization of the pump and collection modes plays a significant role in determining the heralding efficiency, brightness and spectral purity of the source. As such, we first characterized the downconversion source in an unentangled configuration. Bennik provided a comprehensive theoretical description of how the choice of pump and collection modes in collinear sources requires compromises in heralding efficiency, spectral purity and source brightness~\cite{bennink2010optimal}. This theoretical work has since been experimentally supported~\cite{Dixon:2014bg}. To optimize the source heralding efficiency we concede brightness and select a pump waist of $115~\mu\textrm{m}$, and signal and idler waists of $50~\mu\textrm{m}$, yielding a theoretical upper-bound of our heralding efficiency of $96\%$ excluding all optical and detection losses. 

The chosen parameters were implemented with measured raw symmetric heralding efficiencies (we use the definition of Klyshko efficiency~\cite{klyshko80}, i.e. the ratio between \emph{detected} two-fold coincidences over singles of the heralding arm) of up to $(64\pm2)\%$ for both the signal and idler modes , for the $80-200~\mathrm{mW}$ range of laser pump power. Using $80~\mathrm{mW}$ of laser pump power we observed $24300\pm200$ two-fold coincidence counts per second $~\mathrm{c.c./s}$, which corresponds to a pair per pulse generation probability into the collected mode of $p_c\approx0.0007$. Setting pump power below $80~\mathrm{mW}$ started to affect negatively our observed heralding efficiency due to increased contribution from SNSPDs dark counts.  The primary source of loss in this configuration remains the non-unit quantum efficiency of SNSPDs. Straightforward gains can be made to the existing heralding efficiency with better quality optical coatings and background filtering and with improvements in the fiber coupling setup.

\section{Sources of polarization entangled photon pairs}
\begin{figure*}[t]
\begin{center}
\includegraphics[width = 0.95 \textwidth]{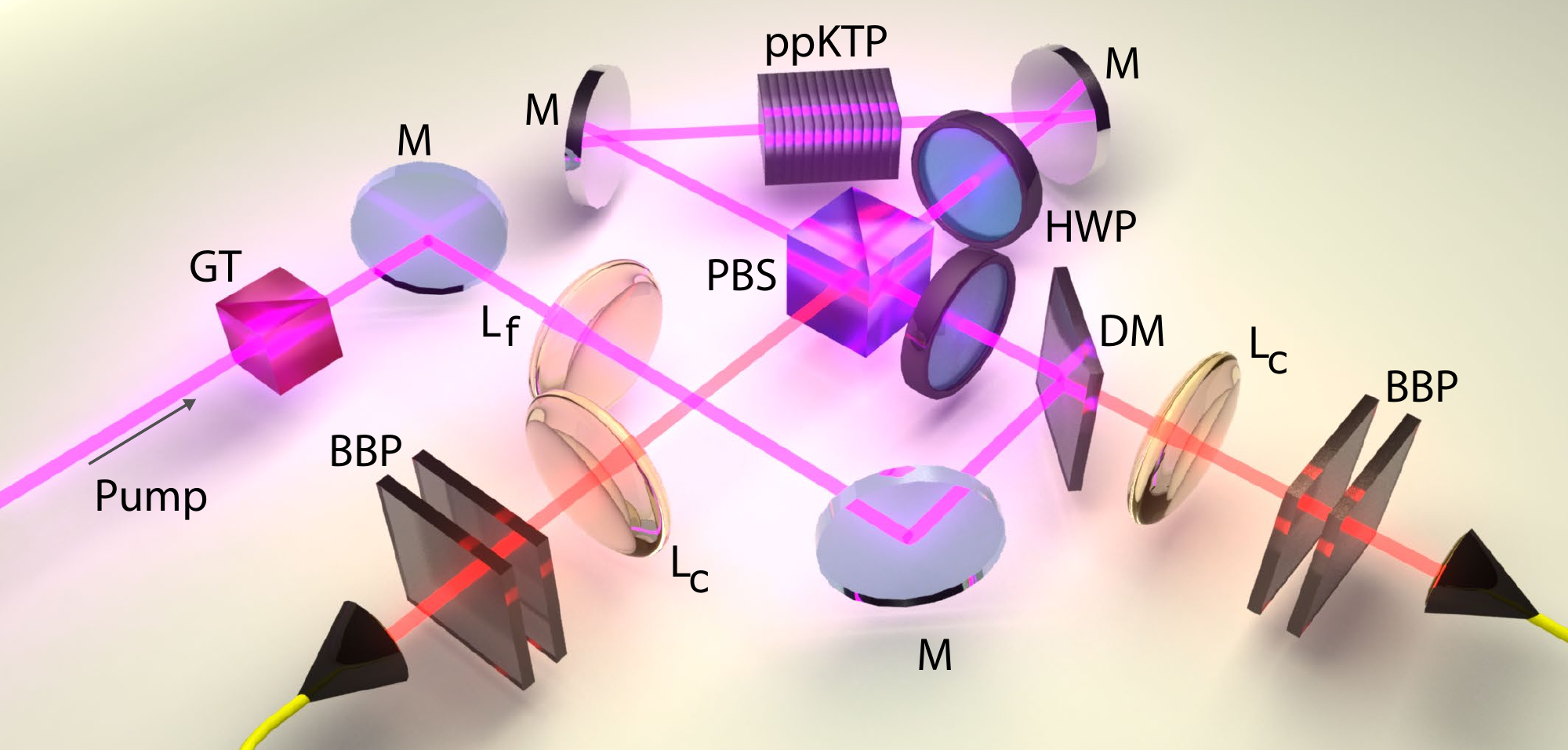}
\caption{Experimental scheme of the polarization-entangled source. A mode-locked Ti:Sapphire laser ($81~\textrm{MHz}$ repetition rate, $785~\textrm{nm}$ wavelength and $\sim \! 170~\textrm{fs}$ pulse length) provides our pump light. The polarizing Sagnac loop comprises a dual-wavelength polarizing beam splitter (PBS), a dual-wavelength half-wave plate (HWP) and a $2~\textrm{mm}$ ppKTP crystal (poling period $46.15~\mu\textrm{m}$, from {\it Raicol}~\cite{disclaimer}). The femtosecond laser pulses are first passed through an optical isolator and Glan-Taylor polarizer (GT) and any residual astigmatism is corrected by a pair of cylindrical lenses. The pump light is focused though a focusing lens ($L_f$) before being guided into the Sagnac loop by mirrors (M) and a dichoric mirror (DM).  The dual-wavelength HWP immediately before the dual-wavelength PBS is set at $22.5^\circ$ for symmetric bi-directional pumping of the crystal.  The internal dual-wavelength HWP is set to $45^{\circ}$. The downconverted photons are collimated through $L_c$ lenses and the pump light and luminescence from optical components is filtered out with $50~\textrm{nm}$ broad bandpass filters (BBP) centered at $1575~\textrm{nm}$. }
\label{experimentalfigure}
\end{center}
\end{figure*}

To polarization-entangle our downconverted photon pair, the aforementioned source is embedded within a polarizing Sagnac interferometer, detailed in Fig.~\ref{experimentalfigure}. In this configuration, a diagonally polarized pump beam is split into its horizontally (transmitted) and vertically (reflected) polarized components on a dual-wavelength polarizing beam splitter (PBS). The two orthogonally polarized pump (labeled as transmitted and reflected from PBS) fields propagate in opposing directions around the Sagnac, bi-directionally pumping the ppKTP crystal centered within the Sagnac loop. 
The  dual-wavelength half-wave plate (HWP) between the reflected port and the crystal serves to  rotate the linear polarization of the reflected pump field to horizontal for downconversion and to rotate the polarization state of the downconverted photons in the transmitted propagating path by 90$^\circ$ to compensate for any temporal walk off. The two counter-propagating downconversion modes are then combined at the dual-wavelength PBS, erasing any distinguishing path information. It should be noted that as the signal (idler) photon is interfered with its counter-propagating signal (idler), the degeneracy of the source should have little effect on the quality of the entanglement. The space constraints of our current-generation Sagnac configuration limited our choice of focusing $L_f$ and collimating $L_c$ lenses, so the pump-collection waists used in the generation of source were less optimal compared to the unentangled configuration.  Using $80~\mathrm{mW}$ of pump power,   we have achieved the symmetric heralding efficiencies of $(45\pm2)\%$  with $45600\pm200~\mathrm{c.c./s}$ and $p_c\approx0.0028$, for the pump waist of $90~\mu\textrm{m}$, and signal and idler waists of $70~\mu\textrm{m}$, which was improved to $(52\pm2)\%$, once the pump waits was set to $100~\mu\textrm{m}$, while the number of observed coincidence counts and pair generation probability reduced to $32900\pm200~\mathrm{c.c./s}$ and $p_c\approx0.0015$ respectively. Other reasons for the reduced efficiency, compared to the unentangled configuration, were the higher number of optical components involved in the setup and increased ellipticity of the pump beam that reduced the mode matching with the optical fiber. We expect higher heralding efficiencies in the future with a new layout design that would account for these imperfections and limitations. 

The femtosecond regime of this Sagnac implementation relaxes a few conditions that are relevant for picosecond and CW implementations. The large bandwidth means the spectral degeneracy of the source is largely insensitive to temperature  fluctuations in the crystal, simplifying two-photon interference between independent sources. Spectral mismatch was suggested as the dominant limitation in a previous demonstrations~\cite{Jin:2015}. As our poling period of $46.15~\mu\textrm{m}$ is phasematched for degenerate downconversion at 1570 nm at $20^\circ C$, our two-photon interference does not require any active temperature stabilization.

\begin{figure}[h]
\includegraphics[width = 0.95 \textwidth]{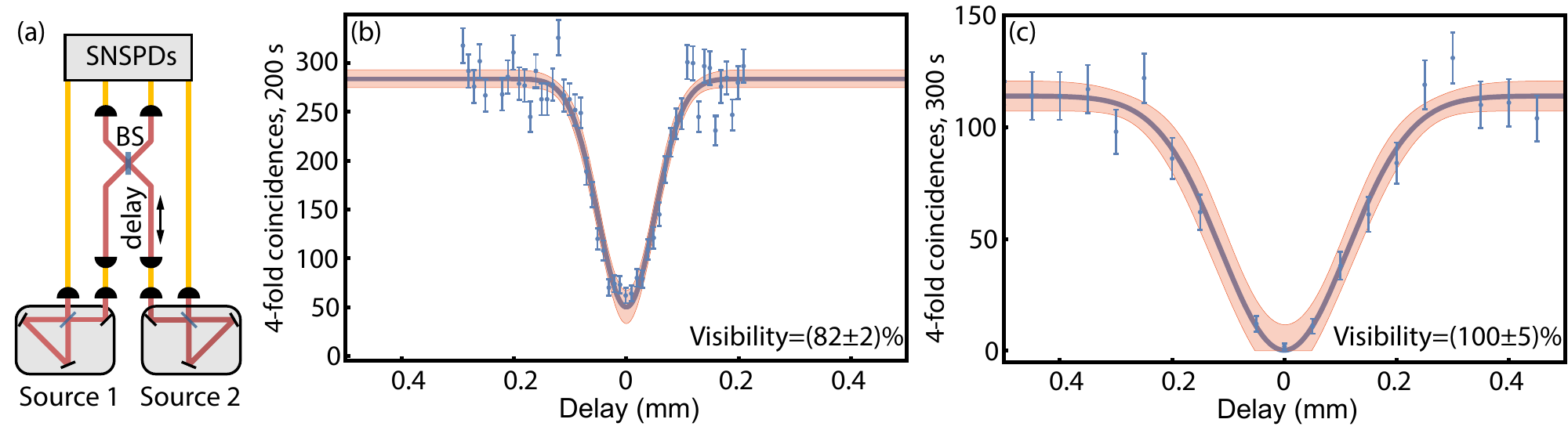}
\caption{ Experimental measurement of two source photon HOM interference. (a) Experimental scheme: heralded photons from two different sources are sent onto a non-polarizing $50:50$ beam splitter (BS) and the temporal distinguishability between the two photons is varied by changing the propagation path length of one of the photons. (b) Measured interference visibility with no filtering applied and (c) with 8 nm bandpass filters in interfering arms only.  The raw visibilities are obtained by a Gaussian fit of the data, without taking into account any limitations due to the beam splitter splitting ratio and without any background subtraction. Shaded areas correspond to a $95\%$ confidence interval for the Gaussian fits, clipped to physically meaningful values.  The visibility uncertainties in (b) and (c) are standard fit errors that arise from the Poissonain statistics of photon counts and interference visibility above $100\%$ is not physically achievable.}
\label{homdips}
\end{figure}

After assembling two Sagnac sources, we have characterized the quality of the produced entangled state.  We used $80~\mathrm{mW}$of pump laser power per source for all the measurements reported below.  Quantum state tomography~\cite{james01} was used to characterize the polarization entanglement of the state. We consider: the fidelity $\mathcal{F} = \bra{\Psi^{-}} \hat{\rho }\ket{\Psi^{-}}$ of the reconstructed two-photon density matrix $\rho$ with the maximally entangled singlet state, $\ket{\Psi ^{-}} = \tfrac{1}{\sqrt{2}}( \ket{H,V} \! - \! \ket{V,H})$; the tangle; and the state purity, as three important parameters~\cite{White:2007} of the state we produced. The measurements  were performed for two cases, as reported in Table 1. 
%~\ref{table:char}. 
For the first case, no filtering was applied, besides the $50~\textrm{nm}$ broad bandpass BBP filters ({\it Alluxa 1575-50 OD4}~\cite{disclaimer}), which were used to block the residual pump and luminescence from optical components and did not affect our two- and four-fold coincidence rates. In second case, we have applied bandpass filters  with a nominal FWHM of $8~\mathrm{nm}$ ({\it Semrock NIR01-1570/3-25}~\cite{disclaimer} ) to our downconverted photons. While the unfiltered case has demonstrated the high quality of our prepared state, we observe further improvement when additional filtering is applied, though at the cost of a reduced heralding efficiency approximately by half. 
\begin{table}[h]\label{table:char}
\centering
\begin{threeparttable}
\caption{Entangled state characterization}
\begin{tabular}{|l|l|l|l|}
\hline
  & Fidelity, \% & Tangle, \% & Purity, \% \\ \hline
BBP filters    & $97.9 \pm 0.1$ & $93.4\pm 0.3$ & $ 96.0\pm 0.2$ \\ \hline
$8~\mathrm{nm}$ filters       & $ 99.0 \pm 0.2$ &  $97.1 \pm 0.4 $ & $98.0 \pm 0.2$\\ \hline
\end{tabular}
\end{threeparttable}
\end{table}
We observe similar results in the quality of our two-source (i.e.\ independent-photon) two-photon Hong-Ou-Mandel visibility measurement, shown in Fig.~\ref{homdips}. Without additional spectral filtering, the theoretical purity of the engineered single-photon source for our chosen parameters is $84\%$ due to the sinc function side lobes of the JSI, as shown in Fig.~\ref{theoryplots}(c). For the unfiltered state we observe a two-photon 4-fold interference visibility of $(82 \pm 2)\%$ (Fig.~\ref{homdips}(b)), which improves to  $(100\pm 5)\%$ with  the addition of the $8~\mathrm{nm}$ bandpass filters in the interfering arms only (Fig.~\ref{homdips}(c)). These visibilities were calculated from raw data without taking into account the imperfect splitting ratio of the $50:50$ (average measured is $\approx48:52$) beam splitter used in the interference setup and without any type of background subtraction. 
For the filtered case, an average interference visibility, calculated from multiple repeated measurements  iterations was found to be $>95\%$, with the drop in visibility being mainly attributed to the temperature variations ($\pm1^{\circ}C$) of the laboratory environment. Once the temperature stability was improved we were repeatedly observing HOM interference with the average visibilities  between $99\%$ and $100\%$.  This is currently the best repeatable two-source photon interference observed that we are aware of for conventional~\cite{kaltenbaek09,tanida12} and telecom group velocity matched~\cite{Jin:2015} sources. While we still require filters to increase our overall source purity beyond the limit of $84\%$, recent work propose directly engineering a Gaussian phase-matching function through sophisticated periodic poling, bringing the purity of the bi-photon state close to unity with only the requirement of background filtering \cite{Branczyk:2011ti,BenDixon:2013tk,Dosseva:2014wr}.  A reason for  significant improvement of interference visibility, over traditional sources, is the presence of the group velocity matching condition which removes one of the two types of time jitter between the independent photon pairs~\cite{tanida12}.  Additionally, the interference visibility is limited by the events when multiple photon pairs are produced from single crystal. For our observed heralding efficiencies and the photon pair production probability of $p_c\approx0.0015$ this upper bound is estimated to be $\approx99.7\%$~\cite{fulconis07}.

These result suggests the large bandwidth of the downconverted photons limits the state quality. An outstanding challenge for our source is the spectral flatness of optical coatings over wide frequency bands. The downconverted photons have a FWHM of approximately  $15~\mathrm{nm}$, which means photons associated with the central lobe of the JSI appreciably exist across a spectral bandwidth of approximately $35~\mathrm{nm}$. With the addition of a $8~\mathrm{nm}$ filter we observe a clear improvement in the quality of our downconverted state. This result supports our main conjecture that the downconverted photons further away from the central wavelength of 1570 nm are partially depolarized by the optical elements within the Sagnac and measurement setup.
It is worth noting, however, that the filtering we apply has a significantly wider pass band than the usual FWHM$\approx 0.5 - 2~\mathrm{nm}$ filters that are used in most multi-photon experiments requiring high state and interference quality~\cite{kaltenbaek09,tanida12}. 

\section{Conclusions}
In conclusion, our results demonstrate the first entangled SPDC source that simultaneously possesses high purity, heralding efficiency and entanglement. With these types of sources, we have observed the highest visibility of non-classical interference between two independent photon sources, to date. While spectral mismatch was suggested as the dominant limitation in a previous demonstrations~\cite{Jin:2014gf}, the femtosecond regime of this Sagnac implementation relaxes a few conditions relevant for picosecond and CW implementations. The large bandwidth of the pump and downconverted photons means the spectral degeneracy of the source is largely insensitive  to small frequency mismatches and phasemathcing temperature fluctuations. Additionally, downconversion at telecom wavelength, where high efficiency SNSPDs perform optimally, makes this source a perfect candidate for long-distance quantum communication experiments.

\section*{Acknowledgments}
Part of this work is supported by ARC grant DP140100648 and part of this work is supported by ARC grant CE110001027. GJP acknowledges an ARC Future Fellowship.

%\bibliography{bib_oe}
%merlin.mbs apsrev4-1.bst 2010-07-25 4.21a (PWD, AO, DPC) hacked
%Control: key (0)
%Control: author (8) initials jnrlst
%Control: editor formatted (1) identically to author
%Control: production of article title (-1) disabled
%Control: page (0) single
%Control: year (1) truncated
%Control: production of eprint (0) enabled
%

\end{document}